Reply to Comment by V. P. Torchigin and A. V. Torchigin [*Phys. Rev. A* **89**, 057801 (2014)] on

"**Theoretical analysis of the force on the end face of a nano-filament exerted by an outgoing light pulse**,"

M. Mansuripur and A.R. Zakharian

[Published in *Physical Review A* **89**, 057802-1~5 (2014).]

**Abstract**. We respond to a Comment on our paper [*Physical Review A* **80**, 023823 (2009)], which appears to have stemmed from a misunderstanding of the various energy-momentum tensors of classical electrodynamics. It is shown that each stress tensor, when used in conjunction with the corresponding force-density and momentum-density expressions, yields results that are consistent with Maxwell's equations and with the conservation laws.

V.P. Torchigin and A.V. Torchigin seem to have a strong preference for one particular energy-momentum tensor of classical electrodynamics to the exclusion of other existing formulations. The Torchigins fault the *Lorentz* formulation for not complying with the requirements of the *Minkowski* formulation. Each formulation has its own stress-energy tensor, electromagnetic (EM) momentum-density, and EM force-density. In Minkowski's case, the EM momentum-density is $\boldsymbol{p}(\boldsymbol{r},t) = \boldsymbol{D} \times \boldsymbol{B}$ and the force-density (in a linear, isotropic, lossless medium) is $\boldsymbol{f}(\boldsymbol{r},t) = -\tfrac{1}{2}\varepsilon_0[\nabla\varepsilon(\boldsymbol{r})]E^2(\boldsymbol{r},t)$. In the case of the Lorentz formulation, the EM momentum-density is the so-called Livens momentum $\boldsymbol{p}(\boldsymbol{r},t) = \varepsilon_0 \boldsymbol{E} \times \boldsymbol{B}$, and the force-density (in non-magnetic media) is given by $\boldsymbol{f}(\boldsymbol{r},t) = (\boldsymbol{P}\cdot\nabla)\boldsymbol{E} + (\partial\boldsymbol{P}/\partial t) \times \boldsymbol{B}$. In what follows we will show the application of each formulation to the examples discussed by the Torchigins. When each formalism is applied correctly and consistently, the results comply with all physical principles and also with known experimental observations.

**Example 1)** The Torchigins' first example pertains to an electrostatic situation. Let a dielectric medium of finite dimensions be subjected to a static electric field $\boldsymbol{E}(\boldsymbol{r})$. Since $\nabla \times \boldsymbol{E}(\boldsymbol{r}) = 0$, we have $\partial E_x/\partial y = \partial E_y/\partial x$, $\partial E_x/\partial z = \partial E_z/\partial x$, and $\partial E_z/\partial y = \partial E_y/\partial z$. Denoting the polarization density of the dielectric medium by $\boldsymbol{P}(\boldsymbol{r}) = \varepsilon_0[\varepsilon(\boldsymbol{r}) - 1]\boldsymbol{E}(\boldsymbol{r})$, the Lorentz formalism yields

$$\boldsymbol{f}(\boldsymbol{r}) = (\boldsymbol{P}\cdot\nabla)\boldsymbol{E} = P_x \frac{\partial \boldsymbol{E}}{\partial x} + P_y \frac{\partial \boldsymbol{E}}{\partial y} + P_z \frac{\partial \boldsymbol{E}}{\partial z}$$

$$= \varepsilon_0[\varepsilon(\boldsymbol{r}) - 1]\left[E_x\left(\frac{\partial E_x}{\partial x}\hat{\boldsymbol{x}} + \frac{\partial E_x}{\partial y}\hat{\boldsymbol{y}} + \frac{\partial E_x}{\partial z}\hat{\boldsymbol{z}}\right) + E_y\left(\frac{\partial E_y}{\partial x}\hat{\boldsymbol{x}} + \frac{\partial E_y}{\partial y}\hat{\boldsymbol{y}} + \frac{\partial E_y}{\partial z}\hat{\boldsymbol{z}}\right) + E_z\left(\frac{\partial E_z}{\partial x}\hat{\boldsymbol{x}} + \frac{\partial E_z}{\partial y}\hat{\boldsymbol{y}} + \frac{\partial E_z}{\partial z}\hat{\boldsymbol{z}}\right)\right]$$

$$= \tfrac{1}{2}\varepsilon_0[\varepsilon(\boldsymbol{r}) - 1]\left[\frac{\partial(E_x^2 + E_y^2 + E_z^2)}{\partial x}\hat{\boldsymbol{x}} + \frac{\partial(E_x^2 + E_y^2 + E_z^2)}{\partial y}\hat{\boldsymbol{y}} + \frac{\partial(E_x^2 + E_y^2 + E_z^2)}{\partial z}\hat{\boldsymbol{z}}\right]$$

$$= \tfrac{1}{2}\varepsilon_0[\varepsilon(\boldsymbol{r}) - 1]\left(\frac{\partial E^2}{\partial x}\hat{\boldsymbol{x}} + \frac{\partial E^2}{\partial y}\hat{\boldsymbol{y}} + \frac{\partial E^2}{\partial z}\hat{\boldsymbol{z}}\right). \tag{1}$$

The force-density of Eq.(1) may now be integrated over the volume of the object under consideration using the method of integration by parts, as follows:

$$\iiint_{-\infty}^{\infty}(\boldsymbol{P}\cdot\nabla)\boldsymbol{E}\,dxdydz = \tfrac{1}{2}\varepsilon_0 \hat{\boldsymbol{x}} \iint_{-\infty}^{\infty}\left[(\varepsilon - 1)E^2 \Big|_{x=-\infty}^{\infty} - \int_{-\infty}^{\infty}\frac{\partial(\varepsilon-1)}{\partial x}E^2 dx\right]dydz$$

$$+ \tfrac{1}{2}\varepsilon_0 \hat{\boldsymbol{y}} \iint_{-\infty}^{\infty}\left[(\varepsilon - 1)E^2 \Big|_{y=-\infty}^{\infty} - \int_{-\infty}^{\infty}\frac{\partial(\varepsilon-1)}{\partial y}E^2 dy\right]dxdz$$

$$+ \tfrac{1}{2}\varepsilon_0 \hat{\boldsymbol{z}} \iint_{-\infty}^{\infty}\left[(\varepsilon - 1)E^2 \Big|_{z=-\infty}^{\infty} - \int_{-\infty}^{\infty}\frac{\partial(\varepsilon-1)}{\partial z}E^2 dz\right]dxdy$$

$$= -\tfrac{1}{2}\varepsilon_0 \iiint_{-\infty}^{\infty}(\nabla\varepsilon)E^2 dxdydz. \tag{2}$$

In the above derivation, the terms involving $[\varepsilon(\mathbf{r}) - 1]\mathbf{E}^2(\mathbf{r})$ at $\pm\infty$ are set to zero because, outside its boundaries, the object is surrounded by vacuum, where $\varepsilon(\pm\infty) = 1$. The *total* force in the Lorentz formalism is thus seen to be identical to that of Minkowski. Therefore, so long as the experimental evidence is based on the *total* force exerted on an *isolated* object, there cannot be any distinction between the electrostatic Lorentz force and its Minkowski counterpart.

It is a well-known fact that different formulations of classical electrodynamics lead to different force-density *distributions* [1,2]. We, among others, have pointed out these differences in several previous publications, and discussed the problem in detail in a recent *Phys. Rev. A* paper [3]. Therefore, to the extent that the Torchigins claim that the force-density *distributions* in the two formulations differ from each other, we do not disagree with them. However, this does not imply the superiority of one theory over the other. The existing experimental evidence (including the nano-fiber experiment of She *et al*, cited by the Torchigins as Ref. [12]) pertains only to *total* EM force exerted on isolated bodies. What is needed is experimental data on force *distribution* within material objects in order to decide among the various EM force formulations.

**Example 2)** In the case of a plane-wave entering at normal incidence from free-space into a dielectric slab of refractive index $n = \sqrt{\varepsilon}$, one must use a light pulse of finite duration in order to see the similarities and differences between the two formulations; see Fig.1. We emphasize that, in *any* analysis of EM systems involving force and torque, the explicit inclusion of the leading and trailing edges of the incident, reflected, and transmitted beams is mandatory. Many controversies and inconsistencies in the published literature stem from neglecting the important property that, in their spatial and temporal extents, all EM waves are finite. From a theoretical standpoint, the necessity of such finite dimensions is dictated by the stress-tensor formulation of electrodynamics, which affirms the conservation laws only when the fields are stipulated to vanish at infinity. From a practical point of view, not only must all conceivable sources of radiation have finite extent, but also they must be turned on at some finite point in time and turned off at a later (finite) time. Therefore, strictly speaking, the conventional assumption that plane-waves extend to infinity in time and space is never justified. Only when the contributions to EM force and torque at the far away "boundaries" of a plane-wave are known to be negligible (or to be irrelevant), can the artificiality of infinite extent be safely retained.

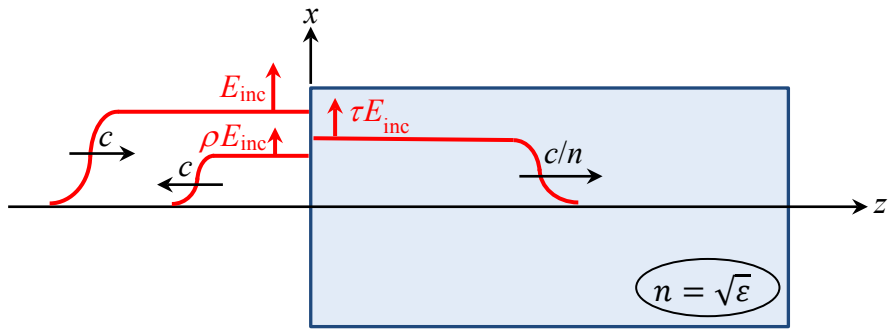

**Fig.1**. A nearly monochromatic plane-wave of finite duration arrives at normal incidence at the front facet of a dispersionless dielectric slab of refractive index *n*. The beam is linearly-polarized along the *x*-axis. The incident, reflected, and transmitted *E*-field amplitudes at the entrance facet ($z = 0$) are $E_{\text{inc}}$, $\rho E_{\text{inc}}$, and $\tau E_{\text{inc}}$, where $\rho = (1-n)/(1+n)$ and $\tau = 2/(1+n)$ are the Fresnel reflection and transmission coefficients of the slab. For each of the incident, reflected, and transmitted waves, the *H*-field is along the *y*-axis, having a magnitude of $E/Z_0$ in vacuum and $nE/Z_0$ within the dielectric; here *E* is the corresponding *E*-field and $Z_0 = (\mu_0/\varepsilon_0)^{1/2}$ is the impedance of free space. The trailing-edge of the incident pulse moves toward the slab at the speed of light in vacuum, *c*, while the leading-edge of the reflected pulse moves away from the slab at the same speed. The leading-edge of the transmitted pulse propagates through the slab at the speed of light $c/n$ within the dielectric.

Returning now to the system depicted in Fig.1, in Minkowski's case, the force density $f(r,t) = -\tfrac{1}{2}\varepsilon_0[\nabla\varepsilon(r)]E^2(r,t)$ acts only on the front facet of the slab, where $\nabla\varepsilon \neq 0$. Since the $E$-field at the interface is $E_x = (1+\rho)E_{\text{inc}}$, where $\rho = (1-n)/(1+n)$ is the Fresnel reflection coefficient and $E_{\text{inc}}$ is the incident field amplitude, the time-averaged Minkowski force per unit-area of the dielectric surface is found to be

$$\langle F_z(z=0,t)\rangle = -\tfrac{1}{4}\varepsilon_0(n^2-1)\left(\tfrac{2}{1+n}\right)^2 E_{\text{inc}}^2 = -\varepsilon_0\left(\tfrac{n-1}{n+1}\right)E_{\text{inc}}^2. \tag{3}$$

The Minkowski force in Eq.(3) is a "pull" force exerted by the EM field at the entrance facet of the slab. Now, the time-rate-of-change of the incident and reflected momenta (per unit cross-sectional area) are given by the speed of light $c$ times the vacuum momentum-density $\wp(r,t) = E\times H/c^2 = (\varepsilon_0 E_x^2/c)\hat{z}$. Combining the contributions of incident and reflected pulses, and also multiplying by ½ (to account for time-averaging), we find

$$\tfrac{d\wp_z}{dt} = -\tfrac{1}{2}\varepsilon_0(1+\rho^2)E_{\text{inc}}^2 = -\varepsilon_0\tfrac{1+n^2}{(1+n)^2}E_{\text{inc}}^2. \tag{4}$$

Inside the dielectric, the $E$-field amplitude is $E_x = \tau E_{\text{inc}}$, where $\tau = 1+\rho = 2/(1+n)$ is the Fresnel transmission coefficient at the front facet. The time-rate-of-change of the EM momentum inside the slab is the velocity $c/n$ of the leading-edge of the light pulse times the Minkowski momentum-density, namely,

$$\tfrac{d\wp}{dt} = \tfrac{1}{2}(c/n)D\times B = \tfrac{1}{2}(c/n)n^2 E\times H/c^2 = \tfrac{1}{2}\varepsilon_0 n^2\tau^2 E_{\text{inc}}^2\,\hat{z} = 2\varepsilon_0\left(\tfrac{n}{1+n}\right)^2 E_{\text{inc}}^2\,\hat{z}. \tag{5}$$

The time-rate-of-change of the *total* EM momentum is obtained by adding Eqs.(4) and (5), that is,

$$\tfrac{d\wp_z}{dt} = 2\varepsilon_0\left(\tfrac{n}{1+n}\right)^2 E_{\text{inc}}^2 - \varepsilon_0\tfrac{1+n^2}{(1+n)^2}E_{\text{inc}}^2 = \varepsilon_0\left(\tfrac{n-1}{n+1}\right)E_{\text{inc}}^2. \tag{6}$$

This is precisely equal in magnitude and opposite in sign to the Minkowski force of Eq.(3), which acts on the entrance facet of the slab. The increase in the total EM momentum of the system given by Eq.(6) is thus balanced by an increase of the mechanical momentum of the slab in the opposite direction; the latter is represented by the time-averaged force $\langle F_z\rangle$ of Eq.(3).

Next, we derive the corresponding results in the Lorentz formulation, where the force exerted on the dielectric is confined to the leading-edge of the transmitted light pulse. In this case $f(r,t) = (P\cdot\nabla)E + (\partial P/\partial t)\times B$. However, $(P\cdot\nabla)E = 0$ for the chosen geometry, and the remaining term yields

$$f(r,t) = \varepsilon_0(\varepsilon-1)\tfrac{\partial E}{\partial t}\times\mu_0 H = \mu_0\left(\tfrac{\varepsilon-1}{\varepsilon}\right)\tfrac{\partial D}{\partial t}\times H = \mu_0\left(\tfrac{n^2-1}{n^2}\right)(\nabla\times H)\times H$$

$$= \mu_0\left(\tfrac{n^2-1}{n^2}\right)\left(-\tfrac{\partial H_y}{\partial z}\hat{x}\right)\times H_y\hat{y} = -\tfrac{1}{2}\mu_0\left(\tfrac{n^2-1}{n^2}\right)\tfrac{\partial H_y^2}{\partial z}\hat{z}. \tag{7}$$

(Note that the force-density at the leading-edge of the pulse does *not* vanish; the argument advanced by the Torchigins based on time-averaging fails when applied to the leading-edge of the pulse.) Integration of the above force-density along the $z$-axis (from $z=0$ to $\infty$), followed by multiplication by ½ (to account for time-averaging), yields the force per unit cross-sectional area of the slab as follows:

$$F_z(t) = -\tfrac{1}{2}\mu_0 \left(\tfrac{n^2-1}{n^2}\right) \int_0^\infty \tfrac{\partial H_y^2}{\partial z} dz = \tfrac{1}{2}\mu_0 \left(\tfrac{n^2-1}{n^2}\right) [H_y(z=0,t)]^2; \tag{8a}$$

$$\langle F_z(t)\rangle = \tfrac{1}{4}\mu_0 \left(\tfrac{n^2-1}{n^2}\right)\left(\tfrac{n\tau E_{\text{inc}}}{\sqrt{\mu_0/\varepsilon_0}}\right)^2 = \varepsilon_0 \left(\tfrac{n^2-1}{n^2}\right)\tfrac{n^2}{(1+n)^2} E_{\text{inc}}^2 = \varepsilon_0 \left(\tfrac{n-1}{n+1}\right) E_{\text{inc}}^2. \tag{8b}$$

Thus, in contrast to the result obtained in the Minkowski case, the Lorentz force exerted on the dielectric (via the leading-edge of the pulse) is seen to be a "push" force.

The time-rate-of-change of the total EM momentum of the system is obtained as before, except that now the *Livens* momentum-density $\boldsymbol{\wp}(\boldsymbol{r},t) = \varepsilon_0 \boldsymbol{E} \times \boldsymbol{B} = \boldsymbol{E} \times \boldsymbol{H}/c^2$ appears inside the slab. Following a similar path that led to Eq.(6), we now find

$$\tfrac{d\wp_z}{dt} = 2\varepsilon_0 \left(\tfrac{1}{1+n}\right)^2 E_{\text{inc}}^2 - \varepsilon_0 \tfrac{1+n^2}{(1+n)^2} E_{\text{inc}}^2 = -\varepsilon_0 \left(\tfrac{n-1}{n+1}\right) E_{\text{inc}}^2. \tag{9}$$

Once again, the time-rate-of-change of the EM momentum of the system given by Eq.(9) is seen to be equal in magnitude and opposite in sign to the net force exerted on the dielectric medium, as given by Eq.(8b).

Both Minkowski and Lorentz formulations thus conserve the total (i.e., EM plus mechanical) momentum. However, the net force, the EM momentum, and the force distribution within the material medium are very different in the two formulations. Moreover, the "push" force predicted by the Lorentz formulation complies with the dictates of the Balazs thought experiment [4], whereas Minkowski's "pull" force does not.

We believe the Balazs thought experiment provides a powerful theoretical argument in favor of the Lorentz formulation and against that of Minkowski. Nevertheless, in the absence of definitive experimental evidence, perhaps one should keep an open mind and allow for the possibility that at least one of the two theories may be incorrect. This, of course, is far from the Torchigins' stance, who seem to believe that every one of the papers published based on the "formula … advanced by Gordon in 1973" [5] is erroneous.

We emphasize that the example cited by the Torchigins involving a semi-infinite dielectric and an infinitely long plane-wave cannot be correctly analyzed unless the situation at infinity is treated with great care. We have chosen a finite-duration pulse of light in the above analysis, precisely to avoid the ambiguities inherent in situations where both the dielectric medium and the light beam have an infinite extent. The Torchigins do *not* allow a leading-edge for the light beam and, therefore, reach the erroneous conclusion that the optical force exerted on the semi-infinite dielectric medium is zero—which would violate momentum conservation. As discussed in the preceding paragraphs, the correct treatment shows that momentum *is* properly conserved. Blind acceptance of infinite extent for a plane-wave has thus led the Torchigins to a fatal error. The inclusion of the leading and trailing edges of the light pulse in the above analysis is *not* optional; the ignorance of this fundamental fact is a major flaw in the Torchigins' argument.

There exist other ways to handle the problems associated with extending the medium and the light beam to infinity along the propagation direction. For example, one might allow for a tiny absorption coefficient in the dielectric medium, so that the incoming light will never reach the far-end of the dielectric slab. Alternatively, one could assume a finite thickness for the dielectric slab, albeit with an antireflection layer placed at the exit facet. Each of these situations can be rigorously analyzed, and the results in each case turn out to be consistent with classical electrodynamics and with the conservation laws.

The important point here is that the Lorentz formalism (based on the application of the Lorentz force law, $\boldsymbol{F} = \rho\boldsymbol{E} + \boldsymbol{J} \times \boldsymbol{B}$, to media that contain electric and/or magnetic dipoles) is a consistent method of calculating EM force and torque exerted on material bodies. This formalism complies with the conservation laws and with the important theoretical argument of Balazs [4]. The Torchigins prefer a different force law (based on Minkowski's stress-energy tensor), and reach different conclusions, which also contradict the Balazs thought experiment [4]. We strongly object to their claim that our method of force calculation based on the Lorentz law is wrong—despite the fact that no experimental evidence has contradicted the predictions of the Lorentz formalism, nor has it been rejected on theoretical grounds involving lack of consistency with well-established physical principles.

The Torchigins' treatment of the Lorentz force, of course, is incorrect, as it leads to a violation of momentum conservation. We have shown here that a correct calculation (i.e., one that incorporates the effects of the leading edge of the light beam within the dielectric medium) removes possible objections (on theoretical grounds) to the application of the Lorentz force law. The question of whether the correct physics is represented by the method of Lorentz or that of Minkowski is an experimental issue which lies outside the domain of the present discussion.

**Example 3)** The Torchigins object to our analysis of a quarter-wave-thick ($\lambda/4$) dielectric slab in conjunction with the Lorentz formulation, citing the violation of Newton's third law (action = reaction) and the existence of the Abraham force. Once again, we believe these objections stem from a misunderstanding of the various formulations of classical electrodynamics. In our calculations of the Lorentz force on a $\lambda/4$-thick slab (described in the Torchigins' Refs. [6,7]), we computed the EM force using the $E$ and $H$ fields of the standing-wave within the slab. This force was then shown to agree with the time-rate-of-change of the overall EM momentum of the system. Similar calculations may be carried out using various other formulations of electrodynamics (e.g., Minkowski, Einstein-Laub, Abraham). In each and every case, the net force exerted on the $\lambda/4$ slab will turn out to be equal in magnitude and opposite in direction to the time-rate-of-change of the total EM momentum of the system.

Momentum continuity is expressed in terms of the EM stress tensor $\overleftrightarrow{\mathcal{T}}$, the EM momentum-density $\boldsymbol{p}$, and the EM force-density $\boldsymbol{f}$, as follows:

$$\overleftrightarrow{\nabla} \cdot \overleftrightarrow{\mathcal{T}}(r,t) + \partial \boldsymbol{p}(r,t)/\partial t + \boldsymbol{f}(r,t) = 0. \qquad (10)$$

In the Minkowski formulation, $\overleftrightarrow{\mathcal{T}}(r,t) = \tfrac{1}{2}(\boldsymbol{D} \cdot \boldsymbol{E} + \boldsymbol{B} \cdot \boldsymbol{H})\overleftrightarrow{\mathbf{I}} - \boldsymbol{D}\boldsymbol{E} - \boldsymbol{B}\boldsymbol{H}$, whereas in the Lorentz formulation $\overleftrightarrow{\mathcal{T}}(r,t) = \tfrac{1}{2}(\varepsilon_o \boldsymbol{E} \cdot \boldsymbol{E} + \mu_o^{-1} \boldsymbol{B} \cdot \boldsymbol{B})\overleftrightarrow{\mathbf{I}} - \varepsilon_o \boldsymbol{E}\boldsymbol{E} - \mu_o^{-1}\boldsymbol{B}\boldsymbol{B}$. Similarly, in the Einstein-Laub formulation $\overleftrightarrow{\mathcal{T}}(r,t) = \tfrac{1}{2}(\varepsilon_o \boldsymbol{E} \cdot \boldsymbol{E} + \mu_o \boldsymbol{H} \cdot \boldsymbol{H})\overleftrightarrow{\mathbf{I}} - \boldsymbol{D}\boldsymbol{E} - \boldsymbol{B}\boldsymbol{H}$. As mentioned earlier, each formulation has its own expressions for EM momentum-density and EM force-density.

In Abraham's formulation, the stress-tensor is the same as that of Minkowski, but the momentum density is $\boldsymbol{p}(r,t) = \boldsymbol{E} \times \boldsymbol{H}/c^2$ rather than $\boldsymbol{D} \times \boldsymbol{B}$. This difference in $\boldsymbol{p}$ results in an additional term, namely, $\partial(\boldsymbol{D} \times \boldsymbol{B} - \boldsymbol{E} \times \boldsymbol{H}/c^2)/\partial t$, in the Abraham force-density formula. In the absence of magnetization, $\boldsymbol{M}(r,t) = 0$, $\boldsymbol{B} = \mu_0 \boldsymbol{H}$, and the extra term becomes $\partial(\boldsymbol{P} \times \boldsymbol{B})/\partial t$, which is the force-density $\boldsymbol{f}_2$ in Eq.(3) of the Torchigins. This is simply the term that must be added to Minkowski's force-density in order to arrive at Abraham's expression for EM force-density. It is therefore not clear what the Torchigins mean when they state that "*This kind of force and the Lorentz force form the Abraham force...*" As far as we know, the Abraham force and the Lorentz force fall into two distinct categories, each with its own expressions of the stress-tensor and the EM momentum-density.

The Torchigins state that "*the momentum of the light wave propagating in a homogeneous optical medium is constant within the optical medium. The same is valid for any number of light* *waves. Thus, there is no change in the momentum of light in a homogeneous optical medium.*" In general, this statement is correct, but the Torchigins proceed to invoke Newton's third law and draw the wrong conclusion from it, as will be explained below.

Interference among various plane-waves propagating inside a homogeneous medium (such as those inside our $\lambda/4$ plate) gives rise to optical fringes where the *E* and *H* fields vary drastically from one place to another. The EM momentum-density, which depends on these fields, thus varies from point to point inside a homogeneous medium. There will be rapid (i.e., at optical frequencies) changes in local EM momentum-density, giving rise to (rapidly varying) local EM forces. However, the time-averaged force-densities arising from such rapid momentum-density fluctuations inevitably vanish. So far we are in agreement with the Torchigins. However, Eq.(10) above indicates that the EM force-density arises not only from the temporal variations of local momentum-density $\boldsymbol{p}$, but also from the divergence of the stress tensor $\overleftrightarrow{\boldsymbol{\mathcal{T}}}$. The interference among two or more plane-waves within a homogeneous medium thus produces spatial variations of the *E* and *H* fields, which lead to spatial variations of the stress tensor. It is these stress-tensor variations (from one location to another inside the homogeneous medium) that produce, in accordance with Eq.(10), the local force-densities inside our $\lambda/4$ plate. This is why the Torchigins' reasoning based on Newton's third law *cannot* apply to EM force and momentum. (As a matter of fact, in the example of the $\lambda/4$ plate, as in any other steady-state situation, the Minkowski force-density is also produced by the divergence of the stress-tensor, *not* by any temporal variations of the local EM momentum-density.)

As a simple example, consider the reflection of a plane-wave at normal incidence from a perfect mirror in vacuum. Obviously, the mirror is pushed by the radiation pressure. However, there is no change of the (time-averaged) EM momentum-density in the homogeneous medium of incidence (vacuum in this case). If the Torchigins' argument were correct, there would be no forces exerted on the mirror, contrary to both theoretical and experimental evidence.

In conclusion, we strongly disagree with the Torchigins' assertion that the Lorentz formalism is a misconception which should be abandoned, and that our analysis (based on the Lorentz formalism) of the nano-fiber experiments of She *et al* has somehow been "erroneous." What the Torchigins have attempted to show is that, under certain circumstances, the predictions of the Lorentz law *differ* from those of the Minkowski theory. However, to the best of our knowledge, none of these differences have been subjected to rigorous experimental verification. Moreover, Minkowski's theory is known to violate the dictates of the Balazs thought experiment [4]. The Torchigins' critique of the Lorentz formalism thus boils down to pointing out certain differences with the predictions of another formalism—that of Minkowski. These are hardly sufficient grounds for rejecting one theory and embracing the other. They do, however, highlight the general areas where experiments are needed to decide which formalism, if either, is correct. We hope that the above explanations have clarified the situation.

1. B. A. Kemp, J. Appl. Phys. **109**, 111101 (2011).
2. W. Frias and A. I. Smolyakov, Phys. Rev. E **85**, 046606 (2012).
3. M. Mansuripur, A. R. Zakharian and E. M. Wright, "Electromagnetic-force distribution inside matter," Phys. Rev. A **88**, 023826~1-13 (2013).
4. N. L. Balazs, "The energy-momentum tensor of the electromagnetic field inside matter," Phys. Rev. **91**, 408-411 (1953).
5. J. P. Gordon, "Radiation forces and momenta in dielectric media," Phys. Rev. A **8**, 14-21 (1973).